\ifx\pdfminorversion\undefined\else\pdfminorversion=7\fi
\ifx\pdfsuppresswarningpagegroup\undefined\else\pdfsuppresswarningpagegroup=1\fi
\documentclass[a4paper,11pt]{article}
\usepackage{pos,multirow,cancel,slashed,ucs,graphicx}
\usepackage[T1]{fontenc}
\graphicspath{{FIGS/}}
\def\color#1{\relax}

\colorlet{yellow}{yellow!70!black}

\newcommand\Dslash{\slashed D}
\DeclareMathOperator\Tr{Tr}
\DeclareMathOperator\Lim{Lim}

\usepackage[capitalize]{cleveref}

\makeatletter
\newcount\Hy@sublinkcounter
\newcommand\phantomsubsection{%
  \Hy@GlobalStepCount\Hy@sublinkcounter
  \xdef\@currentHref{subsection*.\the\Hy@linkcounter.\the\Hy@sublinkcounter}%
  \Hy@raisedlink{\hyper@anchorstart{\@currentHref}\hyper@anchorend}}
\expandafter\renewcommand\expandafter\phantomsection\expandafter
  {\phantomsection\global\Hy@sublinkcounter0\relax}

\@ifundefined{texorpdfstring}
{}{}
\@ifundefined{unichar}
{\newcommand\myunichardef[3]{\expandafter\providecommand\csname text#1\endcsname
                            {#2}}}
{\newcommand\myunichardef[3]{\expandafter\providecommand\csname text#1\endcsname
                            {\unichar{"#3}}}}
\makeatother

\myunichardef{composetilde}{\string~}{0303}
\myunichardef{composemacron}{bar}{0305}
\myunichardef{subscriptthree}{\string_3}{2083}
\myunichardef{superscripttwo}{\string^2}{00B2}

\usepackage{relsize}

\usepackage[acronyms, nohypertypes={acronym}, nopostdot, style=super, nonumberlist, toc]{glossaries}
\glsenableentrycount
\newcommand\ac[1]{\gls{#1}}

\newcommand\acp[1]{\glspl{#1}}

\newacronym{qcd}{QCD}{quantum chromodynamics}
\newacronym{bsm}{BSM}{beyond the standard model}
\newacronym{cp}{CP}{simultaneous interchange of left with right and particle with its antiparticle}
\newacronym{edm}{EDM}{electric dipole moment}
\newacronym{mdm}{EDM}{magnetic dipole moment}
\newacronym{cedm}{cEDM}{chromoelectric dipole moment}
\newacronym{cmdm}{cMDM}{chromomagnetic dipole moment}
\newacronym{qcedm}{qcEDM}{quark cEDM}
\newacronym{qcmdm}{qcMDM}{quark cMDM}
\newacronym{qedm}{qEDM}{quark EDM}
\newacronym{nedm}{nEDM}{neutron EDM}
\newacronym{hisq}{HISQ}{highly improved staggered quark}
\newacronym{CPV}{CPV}{CP-violation}
\newcommand\CPV{$\cancel{\text{CP}}$}
\newacronym{cpv}{\noexpand\CPV}{CP-violating}
\newacronym{awi}{AWI}{axial Ward identity}
\newacronym{esc}{ESC}{excited state contamination}
\newacronym{chipt}{\(\chi\)PT}{chiral perturbation theory}
\newacronym{ckm}{CKM}{Cabbibo-Kobayashi-Maskawa quark-mixing}
\newacronym{mimd}{MIMD}{Multiple Instruction, Multiple Data}
\newacronym{milc}{MILC}{the Multiple Instruction, Multiple Data (MIMD) Lattice Computation}

\renewcommand\subsection[1]{\relax}

\title{Neutron electric dipole moment from isovector quark chromo-electric dipole moment}
\ShortTitle{nEDM from qcEDM}

\author*[a]{Tanmoy Bhattacharya}
\author[b]{Vincenzo Cirigliano}
\author[a]{Rajan Gupta}
\author[a]{Emanuele Mereghetti}
\author[a]{Jun-Sik Yoo}
\author[c]{Boram Yoon}
\affiliation[a]{Los Alamos National Laboratory,\\
  MS B285, P.O. Box 1663, Los Alamos, NM 87545-0285, USA}

\affiliation[b]{Physics Department, University of Washington,\\
  3910 15th Avenue NE, Seattle, WA 98195-1560, USA}

\affiliation[c]{NVIDIA Corporation, Santa Clara, CA 95050, USA}

\emailAdd{tanmoy@lanl.gov}
\emailAdd{cirigv@uw.edu}
\emailAdd{rg@lanl.gov}
\emailAdd{emereghetti@lanl.gov}
\emailAdd{junsik@lanl.gov}
\emailAdd{byoon@nvidia.com}

\abstract{We present results from our lattice QCD study of the
contribution of the isovector \ac{qcedm}
operator to the \ac{nedm}. The
calculation was carried out on four \(2+1+1\)-flavor
\ac{hisq} ensembles (provided to us by \ac{milc}
collaboration~\cite{Bazavov:2010ru,Bazavov:2012xda}) using Wilson-clover quarks to construct correlation
functions. We use the nonsinglet \ac{awi} including
corrections up to \(O(a)\) to show how to control the power-divergent
mixing of the isovector \ac{qcedm} operator with the lower dimensional
pseudoscalar operator.  Results for the \ac{nedm} are presented after
conversion to the \(\overline{\rm MS}\) scheme at the leading-log order.}

\FullConference{The 40th International Symposium on Lattice Field Theory (Lattice 2023)\\
July 31st - August 4th, 2023\\
Fermi National Accelerator Laboratory\\}

\begin{document}
\maketitle
\section{Introduction}
The physics \ac{bsm} of particle physics is needed to explain the observed universe~\cite{Morrissey:2012db}. In particular, such physics needs to violate the symmetry under \ac{cp} to be able to generate the observed excess of matter~\cite{Sakharov:1967dj}. \ac{cpv} interactions can impart  \acp{edm} to nondegenerate quantum eigenstates and observing the \ac{edm} of an elementary particle might be the first indication of such \ac{bsm} physics.

Beyond the scale of electroweak breaking, the standard model can violate \ac{cp} through a gluonic topological term and \ac{CPV} in the couplings of the leptons to the Higgs.  When the weak-symmetry is spontaneously broken, the latter give rise to an irreducible \ac{cpv} phase in the quark mass determinant and four-fermion couplings arising from the single irreducible phase in the \ac{ckm} matrix when the weak gauge bosons are integrated out.  Due to the axial anomaly, in the absence of \ac{bsm} interactions, the \ac{CPV} due to the topological term and that due to the phase of the quark mass determinant can be rotated into each other, and in our previous work, we studied the \ac{nedm} induced by this~\cite{Bhattacharya:2021lol}.  The \ac{nedm} due to the phase of the \ac{ckm} matrix are expected to be much smaller than those due to \ac{bsm} \ac{CPV} in the strong sector.

The lowest mass-dimension \ac{bsm} \ac{cpv} operators are of dimension six.  After weak-symmetry breaking, these give rise (\romannumeral1) to dimension-five \acp{edm} of leptons and both \acp{edm} and \acp{cedm} of quarks, (\romannumeral2) to a dimension-six \ac{cedm} of the gluon, also called the \ac{cpv} Weinberg operator, and (\romannumeral3) to various \ac{cpv} lepton-quark and four-quark four-fermion operators.  The \ac{nedm} due to the \acp{edm} of the quarks are given by the tensor charge, and, in our previous work, we have also calculated these~\cite{Park:2021ypf,Gupta:2018lvp,Jang:2019vkm}.  The corresponding calculations for the \ac{qcedm} and Weinberg operators are preliminary~\cite{Bhattacharya:2023xov}, and no lattice calculation of the \ac{nedm} due to the four-quark operators has been reported yet.

Here we present our recent work~\cite{Bhattacharya:2023qwf} on the \ac{nedm} due to the isovector \ac{qcedm} operator
\begin{equation}
 \bar\psi \Sigma \cdot \widetilde G \tau \psi\,,\label{eq:qcEDM}
\end{equation}
where \(\psi\) denotes the quark-flavor multiplet, \(\tilde G\) the dual chromoelectric field strength, and \(\tau\) a diagonal non-singlet flavor matrix.  This operator is the \(SU(3)\)-color analog of the \ac{qedm} and breaks the chiral symmetry and the discrete symmetries under parity and \ac{cp}, but conserves the charge-conjugation symmetry.  The lattice calculations involving this operator can be conveniently carried out using the Schwinger-source trick: since it is a quark-bilinear, it merely modifies the quark propagator:
\begin{equation}
    {\cal P} = \left[ \Dslash + m - \frac r2 D^2 + c_{\rm SW} \Sigma \cdot G\right]^{-1}
    \rightarrow
    \left[ \Dslash + m - \frac r2 D^2 + \Sigma \cdot \left(  c_{\rm SW} G + i\,\epsilon \tau \widetilde G\right) \right]^{-1}\,,
    \label{eq:Schwinger}
\end{equation}
where \(\cal P\) is the Wilson-clover quark propagator, \(\Dslash\) is the lattice-discretized Dirac operator, \(m\) is the quark mass assumed to be isoscalar, \(r\) is the Wilson parameter, \(c_{\rm SW}\) is the clover parameter, \(\epsilon\) is the strength of the \ac{qcedm} operator, and we have implicitly absorbed powers of the lattice spacing \(a\) to make all quantities dimensionless.  Since the \ac{qcedm} operator is dimension-five, insertion of multiple instances of this operator give rise to uncontrolled \(a^{-1}\) divergences as we take the continuum limit, which necessitates a correspondingly decreasing value for \(\epsilon\).  Details on these issues is presented in our longer publication~\cite{Bhattacharya:2023qwf}.

\begin{figure}
\begin{align}
&\vcenter{\hbox{\includegraphics[width=0.05\textwidth]{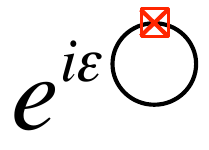}}} \times %{}&\\[3\jot]&
\left(\;\vcenter{\hbox{\includegraphics[width=0.4\textwidth]{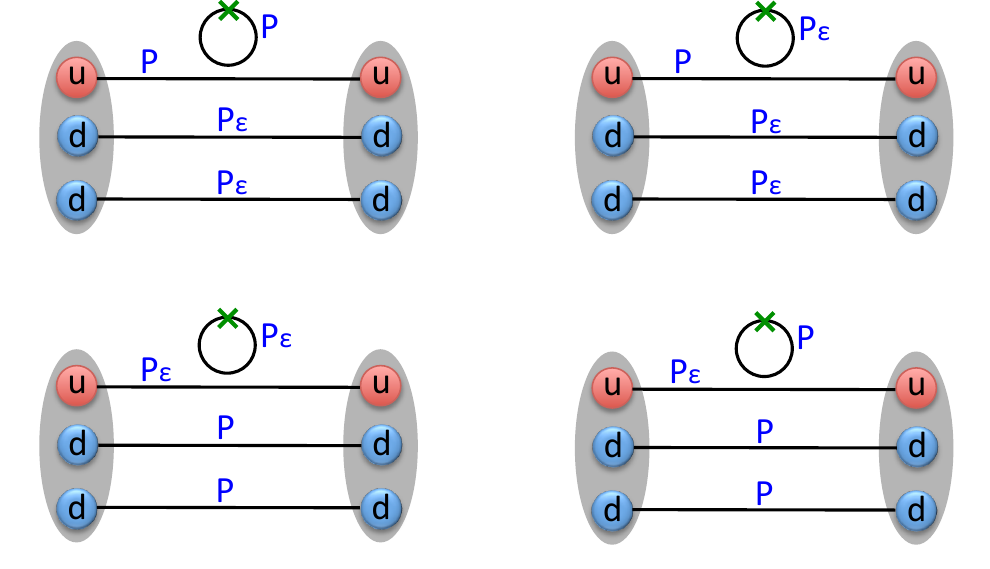}}} + 
 \vcenter{\hbox{\includegraphics[width=0.4\textwidth]{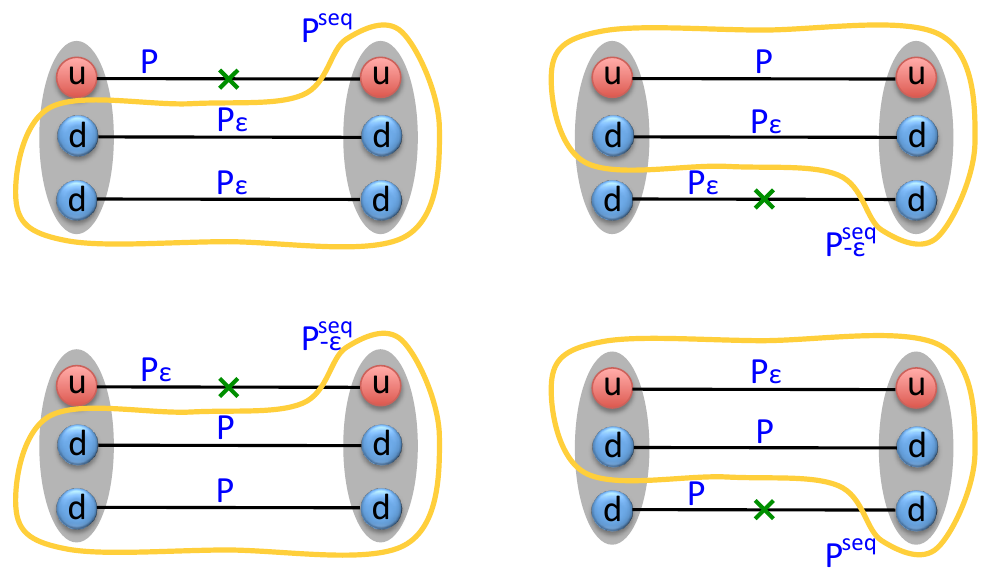}}}\;\right)&\nonumber
\end{align}
\caption{The Wick contractions contributing to the three-point functions}
\label{fig:3pt}
\end{figure}
The \ac{nedm} can be calculated from the two-point function of the nucleon and the three-point function of the vector-current, whose Wick contractions we display in \cref{fig:3pt}.  Since we are concerned with only the isovector \ac{qcedm}, the disconnected loops from the fermion determinant do not contribute, and in this work we ignore the disconnected loops arising from the isoscalar parts of the electromagnetic current, which are found to be small in other matrix element calculations.  As a result, we are left only with connected contributions that we proceed to evaluate.

\begin{table}
{%\small
\setlength{\tabcolsep}{5pt}
\begin{tabular}{ l|lllcr|ll}
 \multicolumn1{c|}{ID} & 
 \multicolumn1c{$a$ (fm)}     & 
 \multicolumn1c{$M^{\rm sea}_\pi$ (MeV)} & 
 \multicolumn1c{$M^{\rm val}_\pi$ (MeV)} & 
 \multicolumn1c{$L^3\times T$}  & 
 \multicolumn1{c|}{$N_{\rm conf}$} &
 \multicolumn1c{\(\epsilon\)} &
 \multicolumn1{c}{\(\epsilon_5\)} \\
\hline 
$a12m310$   & $0.1207(11)$ & $305.3(4)$ & $310.2(2.8)$ & $24^3\times64$  & 1013 & 0.008 & 0.0024 \\
$a12m220L$  & $0.1189(09)$ & $217.0(2)$ & $227.6(1.7)$ & $40^3\times64$  &  475 & 0.001 & 0.0003 \\
$a09m310$   & $0.0888(08)$ & $312.7(6)$ & $313.0(2.8)$ & $32^3\times96$  &  447 & 0.008 & 0.0024 \\
$a06m310$   & $0.0582(04)$ & $319.3(5)$ & $319.3(0.5)$ & $48^3\times144$ &   72 & 0.009 & 0.0012 \\
\end{tabular}}
\caption{The names (ID) and the lattice parameters of the \ac{hisq} ensembles from the \ac{milc} collaboration~\cite{Bazavov:2010ru,Bazavov:2012xda} used in the calculation. \(N_{\rm conf}\) provides the number of configurations analyzed. \(\epsilon\) is defined in \cref{eq:Schwinger} and \(\epsilon_5\) is the corresponding quantity in propagators evaluated with the pseudoscalar operator \(\bar\psi\gamma_5 \tau\psi\) replacing the \ac{qcedm} operator.}
\label{tab:params}
\end{table}
For this calculation, we use a mixed-action setup of tree-level tadpole-improved clover quarks on \ac{hisq} lattices obtained from the \ac{milc} collaboration~\cite{Bazavov:2010ru,Bazavov:2012xda}. All the calculations were done with ensembles where the valence and the sea pion masses were roughly equal, i.e., \(M_\pi^{\rm sea} \approx M_\pi^{\rm val}\); and the lattice sizes in the temporal and spatial directions, \(T \geq L\), were large, \(M_\pi L \gtrsim 4\), where finite volume effects are expected to be small. The lattice parameters are shown in \cref{tab:params}.

\begin{figure}
\begin{center}
  \includegraphics[width=0.34\textwidth]{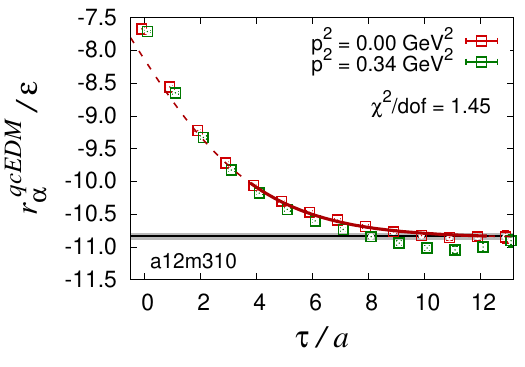}
  \includegraphics[width=0.45\linewidth]{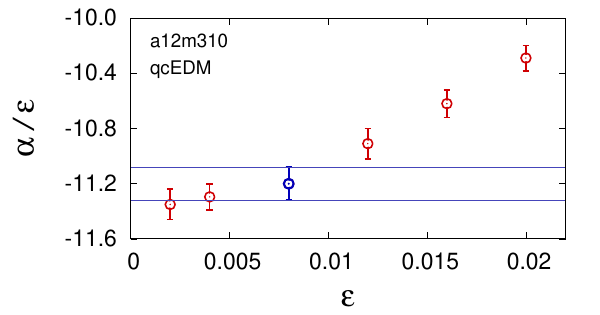}
  \vspace*{-\baselineskip}
  \end{center}  
  \caption{Determination of the \ac{cpv} phase \(\alpha_N\) at various momenta, and checking its linearity.}
  \label{fig:alpha}
\end{figure}
Our operator creating the nucleon in the standard basis~\cite{Pospelov:2000bw,Abramczyk:2017oxr,Bhattacharya:2021lol} is given by \(N_\alpha\) below.  The lattice calculations are actually carried out with \(N_0 \equiv N_{\alpha = 0}\), from which the rotation phase \(\alpha_N\) is determined,
\begin{align}
   N_\alpha &= e^{-i\alpha_N} \epsilon^{abc} \left[\psi_d^{aT} (\gamma_0\gamma_2) \gamma_5 \frac{1\pm\gamma_4}2 \psi_u^b\right] \psi_d^c\\[1\jot]
   \alpha_N &= \Lim_{\tau\to\pm\infty} \frac{\Im \Tr \gamma_5 (1\pm\gamma_4) \langle N_0(0) \bar N_0(\tau)\rangle}{\Re \Tr (1\pm\gamma_4) \langle N_0(0) \bar N_0(\tau)\rangle}%\\&
    \approx -\frac{r\epsilon}{8ma} \frac {a^2\langle \Omega|\bar\psi\Sigma\cdot G\psi|\Omega\rangle}{\langle\Omega|\bar\psi\psi|\Omega\rangle}\,.
\end{align}
where the last expression is a leading order chiral perturbation theory result~\cite{Bhattacharya:2023qwf}. We show an example of the quality of the data determining \(\alpha_N\) in \cref{fig:alpha}, and check that it is linear in \(\epsilon\) and momentum-independent. We then use the determined \(\alpha_N\) to obtain the form factors \(F_{1,2,3}\)
\begin{align}
\langle N_\alpha(p') | J_\mu^{\rm EM} | N_\alpha(p)\rangle %&\\ \span
= \bar u(p') \Big[ \gamma_\mu F_1 + \Sigma_{\mu\nu} \frac {q^\nu}{2M_N} (F_2-i F_3\gamma_5)\Big]u(p)\,,
\end{align}
which decomposition holds when there are no excited state contributions and the theory conserves charge-conjugation.  

\section{Excited State Contribution}

  \begin{figure}
  \leavevmode\hfill\includegraphics[width=0.33\textwidth]{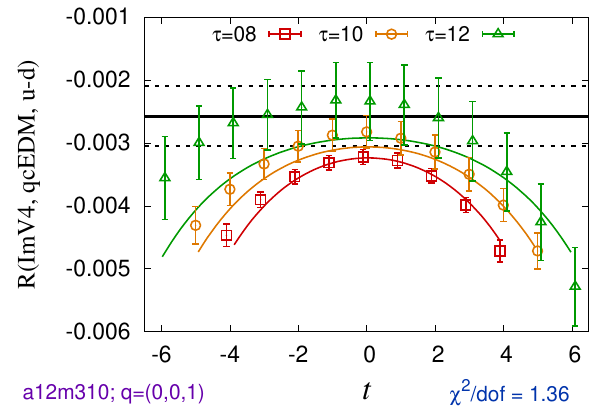}\hfill
  \includegraphics[width=0.33\textwidth]{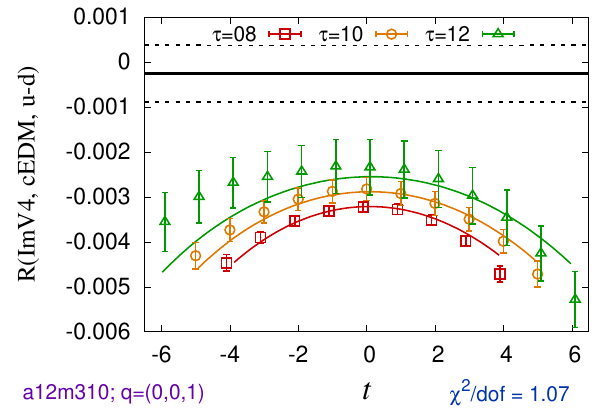}\hfill\hbox{}
  \caption{Effect of excited states on the nucleon matrix elements. The two fits show the same data, but the left hand one assumes a spectrum of excited states obtained from a best fit to the two-point data, whereas the right one assumes a light \(N\pi\) contribution saturating the excited state contribution.}
  \label{fig:esc}
  \end{figure}
The interpolating operators used couple not only to the nucleon state, but other states allowed by the symmetries of the theory. Especially with \ac{CPV}, among these are light multiparticle states like the \(N\pi\) state whose correlations are volume suppressed.  This volume suppression often makes it difficult to see them in two-point correlators.  But since the vector current has a strong coupling to the two pion channel, it is, in principle, possible that the contribution of this state is relatively enhanced in the three-point functions.  Unfortunately, as we found for other matrix elements~\cite{Jang:2019vkm,Gupta:2021ahb,Bhattacharya:2021lol}, a direct check of this is difficult since the \(\chi^2\)-surface has strongly flat directions, and goodness-of-fit tests do not choose between the alternatives. As shown in the example \cref{fig:esc}, this leads to large uncertainties in the final determination of the matrix elements, and, correspondingly, on the predictions for the \ac{nedm}.

\section{Mixing}
Under renormalization, the isovector \ac{qcedm} operator, \(C \equiv C^{(3)}\), has a power divergent mixing with the pseudoscalar operator \(P_3 \equiv P^{(3)} \equiv \bar\psi \gamma_5 \tau \psi\) even when the regularization preserves chiral symmetry.  When chiral symmetry is broken, there is additional divergent mixing with the topological term, which is, however, prohibited for the isovector \ac{qcedm} operator due to the unbroken isospin symmetry in our calculation.

\begin{figure}
\begin{center}
      \includegraphics[width=0.3\textwidth]{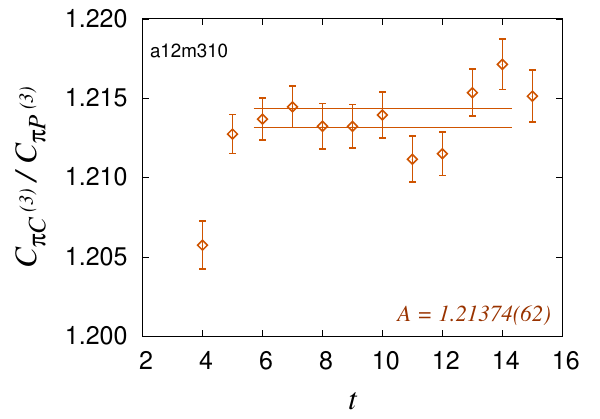}
      \includegraphics[width=0.3\textwidth]{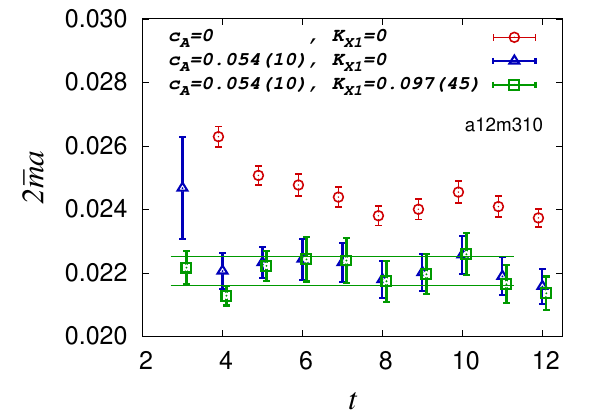}
      \includegraphics[width=0.3\textwidth]{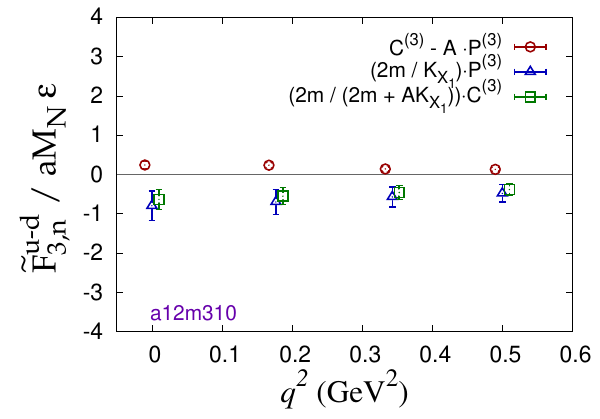}
      \end{center}
      \caption{Example of determination of (left) the constant \(A\) that subtracts the power-divergence of the \ac{qcedm} operator \(C\), (middle) the quark mass \( m\) (labeled \(\bar m\) here) and the constants \(c_A\) and \(K\) (labeled \(K_{X1}\) here) needed for the improved Ward identity in \cref{eq:AWI}, and (right) the \ac{cpv} form-factor \(\tilde F_3\) using three separate lattice operator combinations.}
      \label{fig:F3}
\end{figure}

In the continuum, the isovector pseudoscalar operator can be rotated away by the nonanomalous nonsinglet chiral symmetry, and has no effect.  The lattice situation is more subtle due to the explicit breaking of chiral symmetry.  In fact, the \ac{awi} for Wilson-like fermions is
  \begin{align}
    &\quad Z_A(m) \left[\partial_\mu A_3^\mu + i a c_A \partial^2 P_3 + 2 i m P_3\right]%\\&
    = i a Z_A(m) K \tilde C_3 + O(a^2)
    \label{eq:AWI}
  \end{align}
  where we have restored the explicit powers of the lattice spacing \(a\), \(A_3\) is the axial current, \(\tilde C_3 \equiv C - a^{-2} A P_3\) is defined to be an operator free of power divergence, \(m\) is the quark mass and the term involving \(K\) appears because tree-level tadpole-improved \(c_{\rm SW}\) does not remove all \(O(a)\) effects in the theory even after introducing the improvement constant~\cite{Luscher:1996sc,Bhattacharya:2000pn} \(c_A\) into the axial current.\footnote{For later convenience, we have defined \(c_A\), \(m\) and \(K\) with the axial renormalization constant \(Z_A(m)\) factored out.} Note that there is an \(O(a^2)\) ambiguity in the definition of the coefficient \(A\) that affects only the interpolating operators and not the physical matrix elements, since in the continuum limit a term proportional to \(P_3\) in the action can be rotated away. In our work, we determine by demanding that the vacuum-to-pion matrix element of the operator \(C_3\) is zero at each lattice spacing---since the corresponding matrix element of the interpolating operator \(P_3\) is nonzero, this guarantees the absence of remaining power divergences. This is most conveniently done by taking the ratio of the 2-point correlators of \(C\) and \(P_3\) with a pion interpolating operator \(\pi\) at long Euclidean times (see \cref{fig:F3} left).
  
\begin{table}
{\begin{center}
    \begin{tabular}{|l|l|l|l|l|l|l|}
    \hline
    \multirow{2}{*}{Ensemble}&\multicolumn5{c|}{{\({\tilde F}_3^{P_3}/{\tilde F}_3^{\strut C}\)}}
       &\multirow{2}{*}{\(\displaystyle\frac{K}{2am + A K}\)}\\
       \cline{2-6}
    &\multicolumn1{c|}{\(Q^2=1\)}&\multicolumn1{c|}{\(Q^2=2\)}&\multicolumn1{c|}{\(Q^2=3\)}&\multicolumn1{c|}{\(Q^2=4\)}&\multicolumn1{c|}{\(Q^2=5\)}&\\
    \hline
    a12m310  & 0.879(17) & 0.863(14) & 0.867(18) & 0.844(23) & 0.864(13) & 0.694(48)  \\
    a12m220L & 0.81(10)  & 0.769(77) & 0.869(75) & 0.98(18)  & 0.94(11)  & 0.7807(70) \\
    a09m310  & 1.063(35) & 1.042(40) & 1.078(45) & 1.006(58) & 1.039(44) & 0.740(61)  \\
    a06m310  & &&&&
    & 0.859(64)  \\
    \hline
         \end{tabular}
         \end{center}}
         \caption{Verification of the expected relation between \ac{CPV} due to the pseudoscalar and \ac{qcedm} insertions}
         \label{tab:prop}
         \end{table}
From \cref{eq:AWI}, we immediately find that
      \begin{equation}
        \frac{2am}{K} \frac{P_3}{a} \sim \frac{2am}{2am+K} aC_3 + O(a^2)
        \label{eq:equal}
      \end{equation}
      is power-divergence free, and this quantity gives, up to logarithmic renormalization, the \ac{qcedm} operator in the continuum.  The operator equality of the two sides in \cref{eq:equal} then allows us to determine the constant \(K\) and the quark mass \(m\) (see \cref{fig:F3} middle) and hence the effects of the continuum \ac{qcedm} operator either from the insertion of the lattice pseudoscalar operator \(P_3\), or from that of the lattice \ac{qcedm} operator \(C\). In \cref{tab:prop}, we display the ratio of\footnote{\(\tilde F_3 \equiv F_3 + O(Q^2)\) is a quantity defined~\cite{Bhattacharya:2021lol} and used instead of \(F_3\) because of its better statistical signal.} \(\tilde F_3\) determined from the insertion of either of these two operators along with the value expected from \cref{eq:equal}.  We notice that the expected relation is satisfied to about 10--20\%, which is not surprising since the difference between them, \(O(a^2)/am\), though vanishing in the continuum, could be large for the small quark masses in our calculation.

In addition to determining it from either \(C\), or \(P_3\), we could also determine it directly from \(C_3\).  This, however, involves subtraction of the large power-divergent piece explicitly, and, as shown in \cref{fig:F3} right, shows a result that differs significantly from the other two determinations. Without further understanding of the \(O(a^2)\) errors, it is currently not clear which determination should be preferred.

\section{Renormalization}

After the subtraction of power-law divergences, in our isospin-conserving theory, only logarithmic divergences need to be considered that mix the various operators of dimension-five.  In pure \ac{qcd}, the only mixing is with the \ac{qedm} operator, but this is small, \(O(\alpha_{\rm EM})\sim 1\%\).  Our calculation of \ac{nedm}, however, needs the addition of the electromagnetic interaction \(J^{\rm EM}\cdot A\) to the action.  With this addition, we find that \(\int d^4x \tilde C_3 J^{\rm EM}_\mu A^\mu\) has mixing with \ac{qedm} at \(O(\alpha_s)\).

At leading logarithm (i.e., tree-level matching, one-loop running), we can use this to combine our present results for \ac{qcedm} with our previous analysis of the \ac{qedm} operator:
\begin{subequations}
\begin{align}
  F_3(\vec O_{\overline{\rm MS}}) &= U \setlength{\tabcolsep}{0pt}\begin{pmatrix} \left(\frac{\alpha_s(\mu)}{\alpha_s(a^{-1})}\right)^{-\gamma_{11}/\beta_0}&0\\
    0&\left(\frac{\alpha_s(\mu)}{\alpha_s(a^{-1})}\right)^{-\gamma_{22}/\beta_0}
  \end{pmatrix} U^{-1}\;  F_3(\vec O_L(a))\\
  U &= \begin{pmatrix}1 & -\frac{\gamma_{12}}{\gamma_{11}-\gamma_{22}}\\0&1\end{pmatrix}\qquad\qquad\qquad \vec O = \begin{pmatrix}{\rm \ac{qcedm}}\\{\rm qEDM}\end{pmatrix}\,,
\end{align}
\end{subequations}
where the one-loop \(\beta\)-function and all the one-loop anomalous dimensions \(\gamma\) are known~\cite{Bhattacharya:2023qwf}.

\section{Extrapolation}
The chiral extrapolation of our results needs caution.  \ac{cp}-transformations and chiral rotations do not commute, and the standard CP transformation is to be chosen among a one-parameter family of chirally rotated operators.  Since chiral symmetry is spontaneously broken, the physical CP-violation is the one that leaves the vacuum of the broken theory invariant. The direction of the chiral symmetry breaking is, however, determined by the chiral breaking terms in the theory~\cite{Dashen:1970et}. The analysis of the physical CP-violation in the theory is therefore easiest if we perform a chiral rotation to align the vacuum condensate along the `standard' direction where all the pseudoscalar condensates are zero. In this aligned vacuum, the pions are created by isovector pseudoscalar operators and single pion `tadpoles' are absent.  Starting from a Lagrangian
\begin{equation}
	{\cal L} = {\cal L}^{\rm chiral\hbox{-}\ \&\ CP\hbox{-}conserving} + m \bar\psi \psi + d^{(3)} \bar\psi \Sigma \cdot \tilde G \tau^3 \psi\,,
\end{equation}
the appropriate chiral rotation leads to
\begin{equation}
	{\cal L}^{\rm CPV} = \frac{md_3}{\sqrt{m^2+\bar r^2 d_3^2}} \bar\psi(\Sigma\cdot G - \bar r)\gamma_5 \psi \,,
\end{equation}	
where
\begin{equation}
	2 \bar r \equiv \frac{\langle \Omega | \bar \psi \Sigma \cdot G \psi | \Omega \rangle}{\langle \Omega | \bar \psi \psi | \Omega\rangle}\,.
\end{equation}
This vanishes at \(m=0\), which is consistent with our previous discussion~\cite{Bhattacharya:2015rsa} that when the only chiral violation in a theory is from a single apparently \ac{cpv} operator, the vacuum aligns to maintain the CP symmetry.

\begin{figure}
\begin{center}
\setlength{\tabcolsep}{20pt}
\begin{tabular}{cc}
  \includegraphics[width=0.3\textwidth]{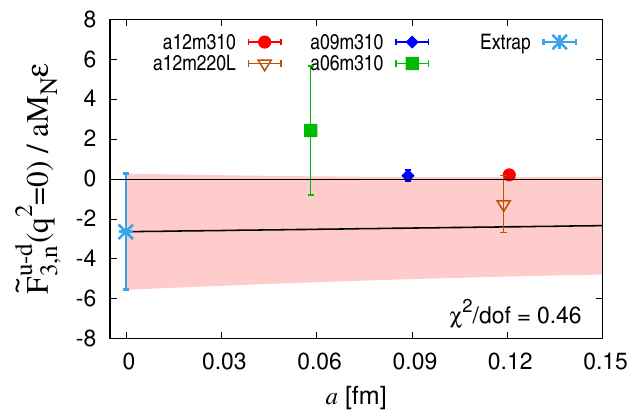} &
  \includegraphics[width=0.3\textwidth]{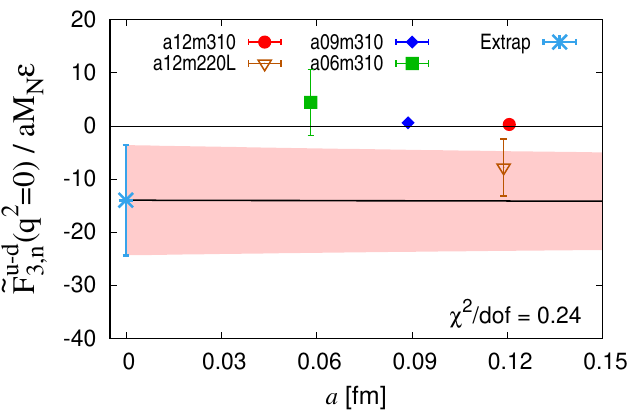}\\
  \includegraphics[width=0.3\textwidth]{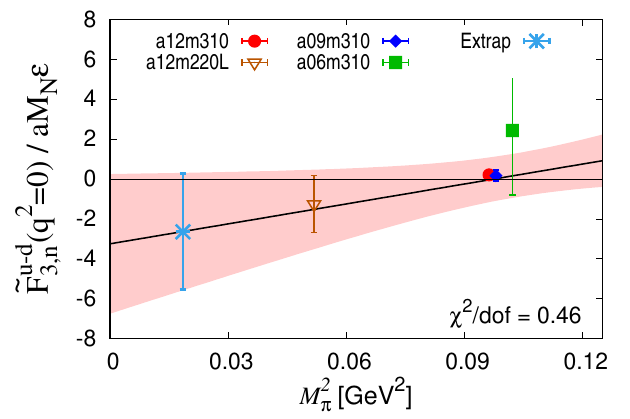} &
  \includegraphics[width=0.3\textwidth]{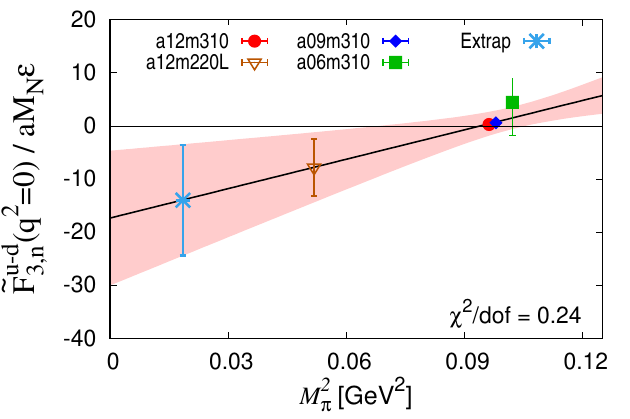}\\
  Standard & \(N\pi\)-fit 
\end{tabular}
\end{center}
\caption{The chiral-continuum extrapolation of the \ac{cpv} form-factor extracted using two choices for the excited state contamination.}
\label{fig:ccfv}
\end{figure}
There is, however, a subtlety in this argument when the \ac{cpv} operator is a power-divergent operator as in our calculation.  In this situation, the operator is ill-defined since the power divergence needs to be controlled. This involves the subtraction of a lower-dimensional operator, and the residual is ill-defined up to finite terms.  One can, in this case, define these finite pieces to set \(\bar r \equiv 0\), and this is the choice we have made with our `subtracted \ac{qcedm}' operator.  It is easy to see that when \(\bar r=0\), the operator and its chiral rotations do not tilt the vacuum manifold, {\it i.e.,} the chiral degeneracy of the vacuum is not lifted by this operator.  In this case, in the absence of any other chiral violation, {\it e.g.,} at the chiral limit in the continuum, chiral dynamics alone does not choose between CP-conserving and \ac{cpv} condensates.  The \ac{CPV} in the chiral-continuum limit of a theory with a standard mass term, with or without the standard Wilson and clover terms, however, chooses the standard orientation of the chiral condensate independent of the mass.  As a result, the \ac{CPV} in this setup persists in the chiral limit.  Thus, in our fits shown in \cref{fig:ccfv}, we do not enforce the vanishing of the \ac{nedm} in the chiral limit.

\section{Conclusions}
In this work we studied the power-divergence of the isovector \ac{qcedm} which is present even with good chiral symmetry. We noticed that the power-divergent mixing is with \(P_3\) which implements chiral rotation, but no CP-violation in the continuum. Since lattice artifacts in this relation are enhanced by \(1/ma\), it is important to demonstrate control. We find that this leads to a large uncertainty when using perturbative \(O(a)\)-improved Wilson fermions, even though the identities following from chiral rotation agree with \ac{chipt} at about 10\%. Finally, we note that control over \ac{esc} still need to be demonstrated.

%\appendix
\acknowledgments
We acknowledge the \ac{milc} collaboration~\cite{Bazavov:2010ru,Bazavov:2012xda} for the lattice ensembles. The calculations used the  CHROMA software suite~\cite{Edwards:2004sx}. Simulations were carried out at (i) the NERSC supported by DOE under Contract No. DE-AC02-05CH11231;  (ii) the Oak Ridge Leadership Computing Facility, which is a DOE Office of Science User Facility supported under Award No. DE-AC05-00OR22725 through the INCITE program project HEP133, (iii) the USQCD Collaboration resources funded by DOE HEP,  and (iv) Institutional Computing at Los Alamos National Laboratory. This work was supported by LANL LDRD program. T.B., R.G. and E.M. were also supported by the DOE HEP and NP under Contract No. DE-AC52-06NA25396. 
V.C. acknowledges support by the U.S. DOE under Grant No. DE-FG02-00ER41132.

\bibliographystyle{JHEP}
\bibliography{main}
\end{document}